\documentclass[letterpaper]{article} 
\usepackage[table]{xcolor}
\usepackage{aaai2026}  
\usepackage{times}  
\usepackage{helvet}  
\usepackage{courier}  
\usepackage[hyphens]{url}  
\usepackage{graphicx} 
\usepackage{natbib}  
\usepackage{caption} 
\usepackage{algorithm}
\usepackage{algorithmic}

\usepackage{newfloat}
\usepackage{listings}
\DeclareCaptionStyle{ruled}{labelfont=normalfont,labelsep=colon,strut=off} 
\floatstyle{ruled}
\newfloat{listing}{tb}{lst}{}
\floatname{listing}{Listing}
\usepackage{xspace}
\usepackage{amsmath, amsfonts}
\usepackage{subcaption}
\usepackage{threeparttable, booktabs, multirow, multicol}
\newcommand{\attack}{{\textsc{\small{CP-Freezer}}}\xspace}
\def\ie{{i.e.},~}
\def\eg{{e.g.},~}
\def\etal{et al.~}
\usepackage{tikz}
\newcommand*\emptycirc[1][0.8ex]{\tikz\draw[thick] (0,0) circle (#1);} 
\newcommand*\halfcirc[1][0.8ex]{%
  \begin{tikzpicture}
  \draw[fill] (0,0)-- (90:#1) arc (90:270:#1) -- cycle ;
  \draw[thick] (0,0) circle (#1);
  \end{tikzpicture}}
\newcommand*\fullcirc[1][0.8ex]{\tikz\fill (0,0) circle (#1);} 
\DeclareRobustCommand*\circled[1]{\tikz[baseline=(char.base)]{\node[shape=circle,draw,color=white,fill=black,inner sep=0.5pt] (char){#1};}}
\newcommand{\Ical}{\mathcal{I}}
\newcommand{\Lcal}{\mathcal{L}}
\newcommand{\Ocal}{\mathcal{O}}
\newcommand{\Pcal}{\mathcal{P}}
\newcommand{\Tcal}{{\mathcal{T}}}
\usepackage{makecell}
\definecolor{Gray}{gray}{0.9}
\definecolor{LightBlue}{RGB}{221,235,247}
\definecolor{LightGreen}{RGB}{230,255,230}

\title{CP-FREEZER: Latency Attacks against Vehicular Cooperative Perception}
\author{
    Chenyi Wang\textsuperscript{\rm 1},
    Ruoyu Song\textsuperscript{\rm 2},
    Raymond Muller\textsuperscript{\rm 2},
    Jean-Philippe Monteuuis\textsuperscript{\rm 3},\\
    Z. Berkay Celik\textsuperscript{\rm 2},
    Jonathan Petit\textsuperscript{\rm 3},
    Ryan Gerdes\textsuperscript{\rm 4},
    Ming Li\textsuperscript{\rm 1}
}
\affiliations{
    \textsuperscript{\rm 1}University of Arizona, 
    \textsuperscript{\rm 2}Purdue University, 
    \textsuperscript{\rm 3}Qualcomm, 
    \textsuperscript{\rm 4}Virginia Tech\\
    \{chenyiw, lim\}@arizona.edu, \{song464, mullerr, zcelik\}@purdue.edu\\ 
    \{jmonteuu, petit\}@qti.qualcomm.com, rgerdes@vt.edu\\
}

\begin{document}

\maketitle


\begin{abstract}
Cooperative perception (CP) enhances 
situational awareness of connected and autonomous vehicles by exchanging and combining messages from multiple agents. 
While prior work has explored adversarial integrity attacks that degrade perceptual accuracy, little is known about CP's robustness against attacks on timeliness (or availability), a safety-critical requirement for autonomous driving. 
In this paper, we present \attack, the first latency attack that maximizes the computation delay of CP algorithms by injecting adversarial perturbation via V2V messages. %
Our attack resolves several unique challenges, including the non-differentiability of point cloud preprocessing, asynchronous knowledge of the victim’s input due to transmission delays, and uses a novel loss function that effectively maximizes the execution time of the CP pipeline. 
Extensive experiments show that \attack increases end-to-end CP latency by over $90\times$, pushing per-frame processing time beyond 3 seconds with a 100\% success rate on our real-world vehicle testbed. 
Our findings reveal a critical threat to the availability of CP systems, highlighting the urgent need for robust defenses.

\end{abstract}


\section{Introduction}\label{sec:intro}

Cooperative perception (CP) allows connected and autonomous vehicles (CAVs) to share sensor data via V2V messages, fusing them to detect objects beyond the line-of-sight and mitigate risks from limited single-vehicle visibility~\cite{cui2022coopernaut}. Despite its promise, CP is vulnerable to adversarial attacks. Previous work has focused on integrity attacks that degrade perception accuracy via object spoofing or removal~\cite{tu2021adversarial, zhang2024datafab}. However, the robustness of CP systems regarding timeliness—a critical \emph{availability} requirement for autonomous driving systems (ADS)~\cite{ADConstraints}—remains unexplored. If a CP system fails to produce timely detections, downstream ADS modules will make uninformed decisions, potentially leading to collisions~\cite{wachenfeld2016worst}. Existing integrity defenses~\cite{amongUs, zhang2024datafab} are ineffective against such threats as they require detection results to be available for their operation.

\begin{figure}[t]
\centering
\includegraphics[width=0.95\columnwidth]{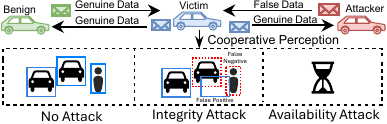}
\caption{Integrity attacks degrade detection quality, whereas \attack \emph{disables} perception by inducing extreme latency, which can lead to severe consequences.}
\label{fig:motivation}
\end{figure}

In this paper, we present \attack, the first latency attack against CP systems. As illustrated in Figure~\ref{fig:motivation}, \attack maximizes the execution time of CP algorithms, disrupting the \emph{availability} of the perception pipeline and endangering real-time vehicle decisions. Realizing \attack involves several challenges:
\underline{C1} \emph{Non-differentiable pre-processing}: Modern CP involves non-differentiable LiDAR feature extraction, complicating gradient-based optimization. Prior solutions for single-agent systems~\cite{liu2023slowlidar} are memory-intensive and too slow for real-time attacks on state-of-the-art (SOTA) point cloud preprocessors~\cite{lang2019pointpillars}.
\underline{C2} \emph{Novel latency-inducing loss for CP}: Previous attacks on CP aim to decrease the detection accuracy regarding particular objects~\cite{zhang2024datafab}, which require precise prior knowledge of the ground truth (otherwise the gradient becomes zero) and cannot increase the execution time of CP. 
Inflating latency poses a fundamentally different objective that needs to be carefully crafted. 
\underline{C3} \emph{Asynchronous attacker knowledge}: In a realistic V2V setting, data is transmitted asynchronously. An attacker cannot rely on a synchronized global state and must craft perturbations with delayed information from the victim.\looseness=-1

We address these challenges as follows. For \underline{C1}, we exploit the trend of intermediate fusion in SOTA CP systems~\cite{xu2022opv2v} by crafting adversarial perturbations at the bird's-eye-view (BEV) feature level, bypassing the non-differentiable steps and reducing the computational burden. For \underline{C2}, our analysis of the CP pipeline reveals that Non-Maximum Suppression (NMS) is a key bottleneck due to its quadratic complexity. We design a novel loss function that maximizes the number of persistent and realistic bounding box proposals to exploit this vulnerability. For \underline{C3}, we use spatial-temporal warping techniques to align the attacker's perturbation with the victim's timestep, compensating for transmission delays and improving attack effectiveness.

Our contributions can be summarized as follows:
\begin{itemize}
    \item We present \attack, the \emph{first} latency attack against CP systems. It exploits feature-level optimization, a novel latency-inducing loss, and spatial-temporal transformation to reveal a critical, overlooked attack vector.
    \item We show that \attack increases end-to-end CP latency by over $90\times$, far surpassing adaptations of existing single-vehicle latency attacks~\cite{ma2024slowtrack,liu2023slowlidar}, and pushes per-frame processing time beyond $3$ seconds with $100\%$ success rates on our real-world testbed.
    \item We provide an extensive evaluation on the benchmark dataset OPV2V~\cite{xu2022opv2v} across multiple hardware platforms and SOTA CP algorithms~\cite{xu2022opv2v, coalign, V2VAM, hu2022where2comm}. Our code is open-source at \url{https://github.com/WiSeR-Lab/CP-FREEZER}.
\end{itemize}


\section{Background and Related Work}
\label{sec:related_work}

\subsubsection{Connected and Autonomous Vehicles (CAVs).} CAVs are vehicles equipped with advanced wireless communication systems, which facilitate real-time data sharing with other connected vehicles (V2V), infrastructure (V2I), and devices (V2X). Their connectivity enhances situational awareness beyond individual vehicles' sensing range and enables collaborative driving. The key technology empowering this is  Cooperative Perception (CP), which extends the perceptual abilities of CAVs by allowing them to exchange and integrate remote sensor information~\cite{f_cooper}. The inference task (\eg 3D object detection) is performed on the collective information fused using the ego vehicle's local data and those received from nearby CAVs.\looseness=-1

\subsubsection{Adversarial Attacks against CP Systems.}
Despite the advantages, CP systems present a novel threat model in which adversaries exploit the data-sharing mechanism~\cite{monteuuis2018attacker}. 
The attacker can inject false data into the CP messages, bypassing physical constraints, making the attack more scalable and easier to execute. 
Previous works have investigated the degradation of detection accuracy and targeted object spoofing/removal by injecting perturbations into shared V2V messages~\cite{tu2021adversarial, zhang2024datafab}. However, these attacks require precise prior knowledge of the ground truth or substantial raw LiDAR data manipulation. Moreover, adversarial defenses for CP have been proposed to validate or restore the integrity of detection results affected by such attacks~\cite{zhang2024datafab, amongUs}.

\begin{table}[t]
\centering
\begin{threeparttable}[t]
\footnotesize
\setlength{\tabcolsep}{3pt}
\begin{tabular}{>{\bfseries}l|c|c|c|c|c}
\toprule
\rowcolor{LightBlue}
Method & \begin{tabular}[c]{@{}c@{}}Attack\\Objective\end{tabular} & \begin{tabular}[c]{@{}c@{}}Real-\\time\end{tabular} & \begin{tabular}[c]{@{}c@{}}Apply to\\SOTA CP\end{tabular} & \begin{tabular}[c]{@{}c@{}}Practical\\Feasibility\end{tabular} & \begin{tabular}[c]{@{}c@{}}Latency\\Effect\end{tabular} \\
\midrule
\rowcolor{Gray}
PGD & Acc. Degrad. & \fullcirc & \emptycirc & \fullcirc & \halfcirc \\
Zhang \etal & \begin{tabular}[c]{@{}c@{}}Spoof/\\Remove Obj.\end{tabular} & \halfcirc & \fullcirc & \halfcirc & \emptycirc \\
\rowcolor{Gray}
SlowTrack & Latency Inc. & \emptycirc & \emptycirc & \emptycirc & \halfcirc \\
SlowLiDAR & Latency Inc. & \emptycirc & \emptycirc & \emptycirc & \halfcirc \\
\rowcolor{LightGreen}
\textbf{Ours} & Latency Inc. & \fullcirc & \fullcirc & \fullcirc & \fullcirc \\
\bottomrule
\end{tabular}
\end{threeparttable}
\caption{Comparison with existing attacks.}
\label{tab:comparison}
\end{table}

\subsubsection{Availability Attacks and Efficiency Robustness.} Recent works in other domains have explored availability as a new attack surface~\cite{wang2022rttee,liu2023slowlidar}. These attacks aim to disrupt the timely and reliable access to information in a system by increasing processing time or energy consumption. Shumailov et al.~\shortcite{shumailov2021sponge} introduced `sponge examples' that maximize the energy and latency of NLP models, while showing marginal impact on vision models. Subsequent works have explored the availability and efficiency robustness of various models, from vision-language models~\cite{gao2024inducing} to dynamic deep neural networks~\cite{Chen_2023_dydnn}. 
In the object detection context, previous works have investigated availability attacks on camera-based 2D object detectors~\cite{shapira2023phantomsponge, ma2024slowtrack} by perturbing the input images, and on LiDAR-based 3D object detection by manipulating the point cloud input~\cite{liu2023slowlidar}.
Nevertheless, availability attacks on safety-critical CP systems have not been explored, which requires new formulations because of their distinct system pipeline, processing protocols, and the unique challenges introduced by the multi-agent collaborative environment.\looseness=-1


\begin{figure}[t]
\centering
\includegraphics[width=\columnwidth]{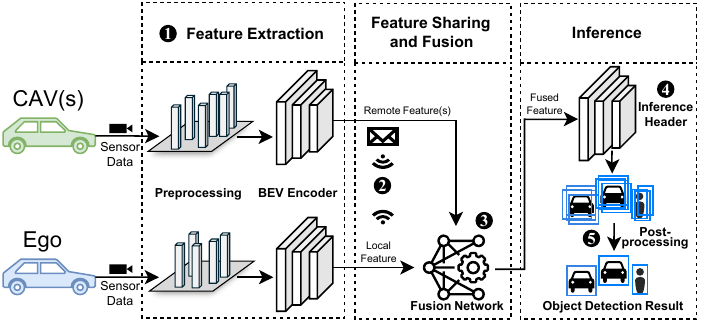}
\caption{Illustration of the intermediate fusion CP pipeline.}
\label{fig:cp_pipeline}
\end{figure}

\section{Methodology}\label{sec:methodology}

\subsection{Availability Attack Surface Analysis}
\label{sec:cp_time_analysis}

\subsubsection{CP Paradigms.} Existing CP algorithms can be categorized into three paradigms based on the stage at which they exchange and fuse data: (1) \emph{Early Fusion}, such as Cooper~\cite{cooper}, entails sharing raw sensor data (\eg LiDAR point clouds) and concatenating the data prior to feature extractions; (2)~\emph{Intermediate Fusion}, exemplified by  AttFusion~\cite{xu2022opv2v} and F-Cooper~\cite{f_cooper}, involves exchanging spatially-aligned BEV features encoded from sensor data, which are then merged using a deep fusion network to form a unified feature for inference; and (3) \emph{Late Fusion}, combines detections from separate vehicles to create an aggregated result~\cite{rauch2012car2x}.\looseness=-1

However, the stringent real-time requirements of ADS~\cite{ADConstraints} pose challenges for early fusion, due to the limited V2X communication bandwidth available (\eg 20 MHz~\cite{v2x_bandwidth}) and prohibitively large volumes of raw sensor data (\eg 4 MB per LiDAR frame~\cite{f_cooper}). In contrast, late fusion suffers from suboptimal detection performance as it solely depends on local information for the inference network~\cite{xu2022opv2v}. Therefore, SOTA CP systems~\cite{xu2022opv2v, coalign, V2VAM, hu2022where2comm} have adopted intermediate fusion, which achieves superior detection accuracy~\cite{opencood} with practical communication and computation overhead. 
Figure~\ref{fig:cp_pipeline} schematically illustrates its core components, and we analyze their potential to be exploited as availability attack surfaces below.

Feature Extraction (\circled{1}) involves local processing of raw LiDAR points into BEV features. The complexity is linear, and the architecture is fixed, offering no viable attack vector.
Message Exchange (\circled{2}) enables an attacker for wireless jamming~\cite {cp_jamming}. However, this would cause the victim to revert to single-vehicle perception, which degrades range but not availability.
Deep Feature Fusion (\circled{3}) module widely adopts dot-product attention~\cite{attention}, which has a complexity of $\Ocal(N^2 \cdot C \cdot H \cdot W)$ for $N$ vehicles. An attack would require deploying many malicious CAVs, which is impractical.
Inference Network (\circled{4}) uses fixed CNN layers, which are not influenced by the attacker.
Post-processing (\circled{5}) is the final step that involves Non-Maximum Suppression (NMS), which has a time complexity of $\Ocal(M^2)$ for $M$ proposals~\cite{NMS}. Therefore, by adversarially inflating the number of high-confidence, persistent proposals $M$, we can trigger a quadratic blow-up in processing time, paralyzing the CP system.\looseness=-1

\subsection{Problem Formulation}

\subsubsection{System Model.}

In a multi-vehicle CP system with $N$ CAVs, each vehicle $V_i$ computes a BEV feature $X_i^t$ at time $t$. A victim $V_v$ fuses its feature with those from neighbors $\mathcal{V}_v$ using a fusion network $\mathcal{F}(\cdot)$ to get $\tilde{X}_v^t$. This is fed to an inference head $\mathcal{I}(\cdot)$ to generate proposals $\mathcal{P}_v^t$, which are finalized by $\text{NMS}(\cdot)$.

\subsubsection{Threat Model.}

We assume a white-box attacker controlling a compromised CAV. The attacker can manipulate its outgoing V2V messages to maximize the victim's end-to-end latency $\Tcal(\cdot)$. The goal is to find a perturbation $\delta^t$ that solves:
$\delta^t=\arg\max_{\delta}\Tcal\left(\text{NMS}\Bigl(\Ical(\tilde{X}_v^t )\Bigr) \right)$, where $\tilde{X}_v^t$ is the fused feature map containing the attacker's perturbed feature $X_a^t+\delta$.

\begin{figure}[t]
\centering
\includegraphics[width=\columnwidth]{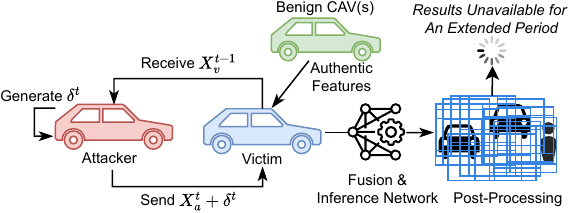}
\caption{Overview of \attack.}
\label{fig:attack_system}
\end{figure}

\subsubsection{Attack Overview.}
As shown in Figure \ref{fig:attack_system}, \attack generates perturbations at the feature level, bypassing non-differentiable preprocessing. The attacker optimizes a perturbation using our latency-inducing loss function and injects it into its shared feature. To handle asynchronous communication, where the attacker only has the victim's feature from $t-1$ when crafting an attack for time $t$, we use a feature warping technique to predict the victim's state and enhance attack effectiveness.\looseness=-1

\subsection{Adversarial Loss Function Design}
\label{sec:loss_design}

To exploit the quadratic time complexity of NMS, the attacker needs to inflate the number of pre-NMS bounding box proposals while ensuring that the resulting proposals remain physically plausible and sparsely distributed to resist efficient filtering. To this end, we combine multiple loss terms that work in tandem: 
$(1)$ \emph{Confidence Activation}, to produce high-confidence proposals at as many spatial locations as possible; 
$(2)$ \emph{Shape Regularization}, to prevent the boxes from becoming abnormally large and reduce overlaps; 
and $(3)$ \emph{Vertical Plausibility}, to ensure that the proposals are within a realistic vertical range (\eg the ground plane to the height of the car roof). 
In the following, we outline each term and its interplay in the final joint loss.

\subsubsection{Confidence Activation Loss. }
We design a loss to increase the CP model's objectiveness scores at every anchor box to enable as many high-confidence proposals to be produced as possible. 
However, pushing scores arbitrarily high wastes gradient steps that could be spent elevating other locations that have not yet passed the confidence filtering threshold. 
Therefore, we include an activation control parameter \(\tau\), chosen to match the pre-NMS filtering threshold. 
Formally, let \({\mathbf{s}} \in \mathbb{R}^{B\cdot H\cdot W}\) be the predicted objectiveness scores for the anchor boxes\footnote{$B$ is the number of anchor boxes at each feature location. $H,W$ are the height and width of the feature map, respectively.}, then the confidence activation loss is $\Lcal_{\text{conf}}=\frac{1}{BHW}\left({\mathbf{1}}^\top\text{ReLU}(\tau\mathbf{1}-{\mathbf{s}})\right)$, 
where ${\mathbf{1}}=[1,...,1]\in \mathbb{R}^{B\cdot H\cdot W}$. Minimizing \(\mathcal{L}_{\text{conf}}\) effectively pushes confidence of potential proposals to at least \(\tau\), but does not spend additional optimization effort to push them beyond \(\tau\). This mechanism makes the attack more efficient by distributing gradient resources to those boxes that still lie below the threshold, thereby maximizing the total count of high-confidence proposals entering NMS.

\subsubsection{Shape Regularization Loss. }
Simply inflating confidence scores could yield proposals oddly shaped (excessively large) or overlap each other excessively. Consequently, these fabricated proposals can be efficiently pruned by sanity checks and NMS in CP postprocessing~\cite{opencood}, undermining the goal of maximizing latency. 
To address this, let $M$ be the total number of 3D bounding box proposals and let each proposal be parameterized by 
$\mathbf{b}_i = (x_j, y_j, z_j, \ell_j, w_j, h_j,\sigma_j)$\footnote{$(x_j, y_j, z_j)$ are the center coordinates of the proposal. \(\ell_j, w_j, h_j\) denote the length, width and height of the proposal, and $\sigma_j$ is the yaw.}.  
We impose the following shape regularization loss $    \mathcal{L}_{\text{shape}}
    \;=\;
    \frac{1}{M}\sum_{j=1}^M 
    \mathbf{I}\!\Bigl[
        \ell_j > L_{\max} 
        \;\lor\; 
        w_j > W_{\max}
    \Bigr],$ 
where \(\mathbf{I}[\cdot]\) is an indicator function, and \(L_{\max}, W_{\max}\) define upper bounds on length and width (\eg 5\,m). By discouraging large boxes, we reduce the IoU between proposal pairs so that they are more sparsely distributed. As more proposals survive each iteration of NMS, the total number of NMS comparisons (\(\mathcal{O}(M^2)\)) and the overall latency are inflated. 

\subsubsection{Vertical Plausibility Loss. }
To further maintain realism in 3D placement, we penalize bounding boxes whose vertical coordinate \(z_j\) is implausible (\eg too high above the road or deep below ground). Let \(\mathcal{Z} = [z_{\min}, z_{\max}]\) be a valid elevation range (\eg \([1, 3]\) meters). We define $    \mathcal{L}_{\text{vertical}} 
    \;=\; \frac{1}{M}\sum_{j=1}^M \mathbf{I}[\,z_j < z_{\min} \;\lor\; z_j > z_{\max}\,].$ 
Penalizing out-of-range boxes deters trivial outliers that CP algorithm implementations typically reject by domain logic or height-thresholding~\cite{opencood}.\looseness=-1

\subsubsection{Unified Loss Formulation. }
We combine the above terms into a single objective that the attacker minimizes, with hyperparameters \(\lambda_1\) and \(\lambda_2\) balancing the relative emphasis $\Lcal_{\text{total}}=\lambda_1\Lcal_{\text{score}}+\lambda_2(\Lcal_{\text{shape}}+\Lcal_{\text{vertical}})$, 
Minimizing this adversarial loss increases the number of high-confidence, plausible, and persistent proposals. Consequently, a large fraction of these proposals $(1)$ survive the initial confidence-based filtering, $(2)$ are difficult for standard sanity checks to block preemptively, and $(3)$ cannot be efficiently eliminated during NMS.

\subsection{Spatial-Temporal Feature Warping}
\label{sec:feature_warping}

To perform online and real-time optimization for the adversarial perturbation to effectively increase the victim's CP latency, the attack needs to account for delayed information in a realistic CAV environment. In other words, the attacker must generate perturbations for the current frame \(t\) with access only to the victim's feature from \(t-1\) due to asynchronous transmission.
To overcome this challenge, we exploit the spatial-temporal consistency property and warping techniques of BEV features, 
where consecutive feature maps exhibit geometric shifts corresponding to ego-motion and can be approximated by transformation~\cite{yu2024feature_flow}. 

\subsubsection{Affine Transformation Derivation. } 
Let \(\mathbf{T}^{t-1 \rightarrow t}_v\in \mathbb{R}^{3\times 3}\) be the 2D affine transform from the victim’s pose at \(t-1\) to its pose at \(t\). We derive \(\mathbf{T}^{t-1 \rightarrow t}_v\) by comparing the victim’s reported positions \(\mathbf{p}^{t-1}_v\) and \(\mathbf{p}^{t}_v\) through standard pose subtraction~\cite{yu2024feature_flow, coalign}. Similarly, the adversary estimates \(\mathbf{T}^{t-1 \rightarrow t}_a\) for its own motion. These transformations are then normalized according to the BEV resolution, mapping physical distances to feature pixels (\eg 0.4\,m/grid cell).\looseness=-1

\subsubsection{Warping and Online Optimization. }
Given a prior feature map \({X}^{t-1}_v\in \mathbb{R}^{C\times H\times W}\), we warp it into the new frame \(t\) via $    \widetilde{{X}}^{t}_v 
    \;=\;
    \mathcal{W}\!\bigl({X}^{t-1}_v,\;{T}^{t-1 \rightarrow t}_v\bigr),$ 
where \(\mathcal{W}(\cdot,\cdot)\) denotes a spatial transformation that applies \(\mathbf{T}^{t-1 \rightarrow t}_v\) to align the feature map with the updated coordinate frame. The adversary performs the same operation for its own feature \(\mathbf{X}^{t-1}_a\), resulting in \(\widetilde{\mathbf{X}}^{t}_a\), which is used to optimize the adversarial perturbation.


\section{Evaluation}\label{sec:eval}

\subsection{Experimental Setup}

\subsubsection{Dataset and Algorithms.}  We evaluate \attack on the widely used benchmark CP dataset OPV2V~\cite{xu2022opv2v}, which is a large-scale V2V perception dataset comprising 70 varied scenes across 8 towns within the digital twin simulator CARLA~\cite{Dosovitskiy17Carla}. 
The dataset contains 11,464 LiDAR point cloud frames from 2-5 CAV agents and includes 232,913 annotated 3D vehicle bounding boxes. 
We regard the CAVs with the lowest and second-lowest indices as the victim and attacker vehicles, respectively. 
We assess \attack on four SOTA CP algorithms: AttFusion~\cite{xu2022opv2v}, Where2comm (W2C)~\cite{hu2022where2comm}, CoAlign~\cite{coalign}, and V2VAM~\cite{V2VAM}, which are known for their high detection accuracy and real-time computational efficiency~\cite{opencood}.\looseness=-1 

\subsubsection{Comparison Baselines.}
As the first CP latency attack, we compare \attack against two baselines that are both optimized at the feature level for a fair comparison:
$(1)$ Projected Gradient Descent (PGD)~\cite{MadryMSTV17PGD}, which though primarily aims to degrade accuracy untargetedly, often inflates the number of false positives as a side effect, thereby contributing to post-processing overhead. 
$(2)$ Prior art in single-vehicle latency attacks proposed by SlowLiDAR~\cite{liu2023slowlidar} and SlowTrack~\cite{ma2024slowtrack}. We use their loss function and adapt to CP with multi-agent input for comparison.

\subsubsection{Evaluation Metrics.}
\begin{figure*}[t]
\centering
    \begin{tabular}[t]{c}
    \centering
    \begin{subfigure}[t]{0.32\textwidth}
        \includegraphics[width=\linewidth]{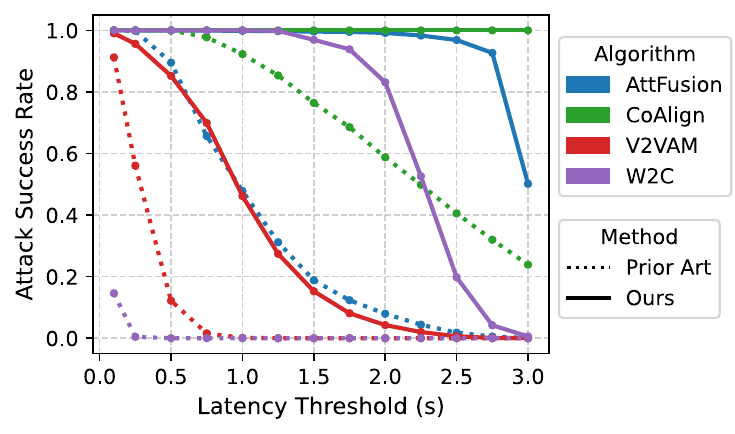}
        \caption{RTX 2080 Super}
        \label{fig:asr_2080}
    \end{subfigure}
    \begin{subfigure}[t]{0.32\textwidth}
        \includegraphics[width=\linewidth]{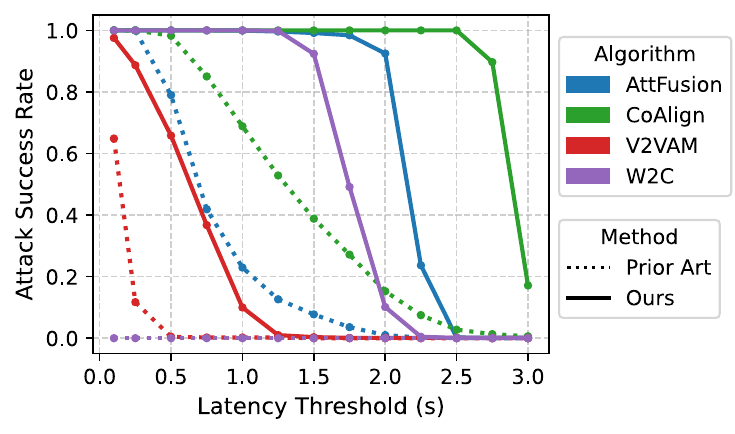}
        \caption{RTX 3060 Ti}
        \label{fig:asr_3060}
    \end{subfigure}
    \begin{subfigure}[t]{0.32\textwidth}
        \includegraphics[width=\linewidth]{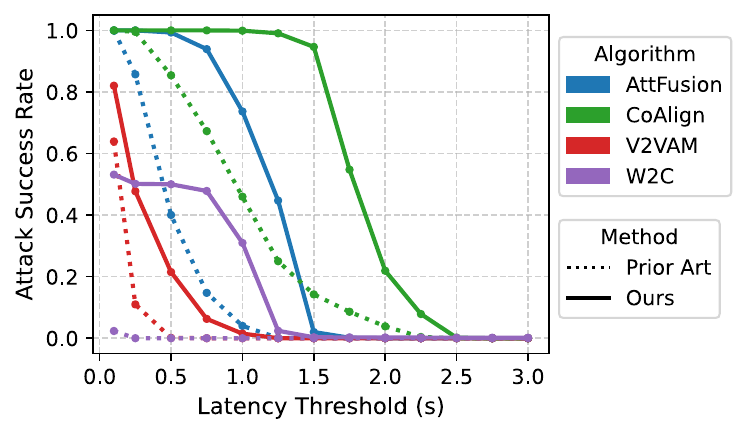}
        \caption{RTX 4090}
        \label{fig:asr_4090}
    \end{subfigure}
\end{tabular}
\caption{Attack success rates (ASRs) under varying latency thresholds, across different hardware. For each CP algorithm, \attack achieves higher ASRs compared to the attack adopting the adversarial loss function used in the prior art on single-vehicle latency attacks.}\label{fig:asr}
\end{figure*}

We measure the average end-to-end computation latency of the victim’s CP pipeline, both in benign and attacked scenarios (\ie $\bar{T}_{\text{benign}}$ and $\bar{T}_{\text{attack}}$). 
To measure the variability of the CP latency, we use relative standard deviation $(\%\text{RSD})$ defined as the ratio of the latency standard deviation over the average latency. A lower $\%\text{RSD}$ indicates that the measured latency values concentrate more consistently around the mean. 
Also, since the execution time is hardware-dependent, we calculate the Rate-of-Increase of pre-NMS proposal counts (RoI-P) and latency of the whole CP pipeline (RoI-L) defined as $\text{RoI-}{\text{L}}=\frac{\bigl({T}_{\text{attack}} - {T}_{\text{benign}}\bigr)}{{T}_{\text{benign}}},$ and $\text{RoI-}{\text{P}}=\frac{\bigl({|\Pcal|}_{\text{attack}} - {|\Pcal|}_{\text{benign}}\bigr)}{{|\Pcal|}_{\text{benign}}}$. 
The RoIs provide hardware-independent measurements that capture how much slower the CP system runs and the increase in the number of bounding box proposals when attacked, relative to the benign scenario~\cite{liu2023slowlidar, ma2024slowtrack}.
Furthermore, to assess whether the increased latency poses a significant safety risk, we consider an attack as successful if it pushes the end-to-end CP latency beyond the threshold of \(1.5\) seconds, which is widely considered the worst-case time-to-collision threshold in autonomous driving literature~\cite{wachenfeld2016worst, kusano2011sae_method}. The attack success rate (ASR) is the ratio of successful attacks to the total number of frames. Also, we show ASR under different latency thresholds in Figure~\ref{fig:asr}.

\begin{table}[t]
\centering
\begin{threeparttable}[t]
\footnotesize
\setlength{\tabcolsep}{3pt} 
\begin{tabular}{l|c|c|c|c}
\toprule
\rowcolor{LightBlue}
Model & No Attack & PGD & Prior Art & \textsc{CP-Freezer} \\
\midrule
AttFusion & 44.02 & \makecell{321.64\\(12.28$\times$)} & \makecell{432.92\\(13.26$\times$)} & \makecell{\textbf{1235.53}\\(\textbf{46.19$\times$})} \\ \hline
CoAlign & 46.52 & \makecell{446.24\\(12.93$\times$)} & \makecell{626.36\\(17.41$\times$)} & \makecell{\textbf{1602.80}\\(\textbf{54.84$\times$})} \\ \hline
V2VAM & 21.07 & \makecell{52.32\\(3.70$\times$)} & \makecell{171.06\\(9.58$\times$)} & \makecell{\textbf{340.46}\\(\textbf{21.09$\times$})} \\ \hline
Where2comm & 19.44 & \makecell{5.58\\($-$0.62$\times$)} & \makecell{45.53\\(2.02$\times$)} & \makecell{\textbf{506.20}\\(\textbf{37.81$\times$})} \\ 
\bottomrule
\end{tabular}
\begin{tablenotes}[flushleft]\footnotesize
\item[$*$] Average RoI is computed per frame and then averaged. It differs from the RoI of $\bar{|\Pcal|}$.
\end{tablenotes}
\end{threeparttable}
\caption{\attack effectively inflates NMS workload. It has the highest average number of pre-NMS proposals $\bar{|\Pcal|}$ and RoI in proposals $\bar{\text{RoI-}}\text{P}$ (parentheses).}
\label{tab:roi_proposal}
\end{table}

\subsubsection{Testing Hardware.}
We evaluate \attack 
across three different hardware platforms: 
$(1)$ RTX 2080 Super: a high-performance GPU that is equipped on our real-world vehicle testbed running a NUVO-8208GC computer. We emulate sensor inputs and V2V communication by loading the dataset through the vehicle developer I/O. 
$(2)$ RTX 3060 Ti: a more recent GPU with better performance-per-watt efficiency than RTX 2000 series. 
$(3)$ RTX 4090: a top-tier GPU, representing a high-end system with ample computing power and memory bandwidth.

\begin{figure}[t]
\centering
\includegraphics[width=\columnwidth]{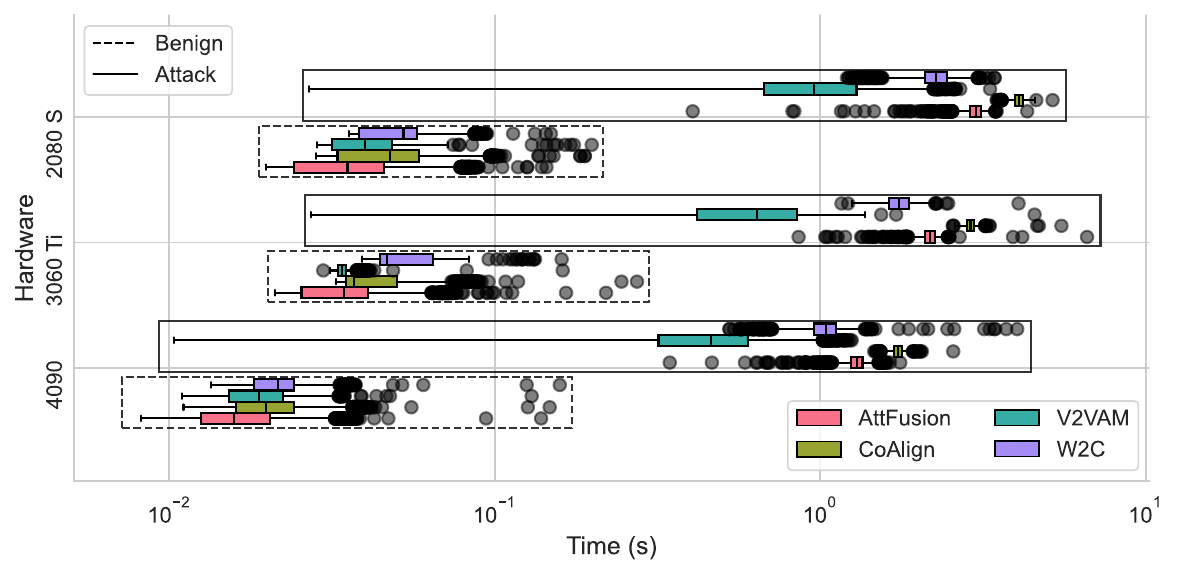}
\caption{Latency distribution boxplot. Each CP algorithm mostly runs below 100\,ms without attack. When \attack is applied, the latency increases and concentrates around several seconds.}
\label{fig:latency_boxplot}
\end{figure}

\begin{table*}[t]
\centering
\setlength{\tabcolsep}{2pt}
\footnotesize
\begin{tabular}{l|l|cc|ccc|ccc|>{\columncolor{LightGreen}}c>{\columncolor{LightGreen}}c>{\columncolor{LightGreen}}c}
\toprule
\rowcolor{LightBlue}
\multicolumn{2}{c|}{Attack} & \multicolumn{2}{c|}{No Attack} & \multicolumn{3}{c|}{PGD} & \multicolumn{3}{c|}{Prior Art} & \multicolumn{3}{c}{\attack} \\
\midrule
Hardware & Model & Avg&\%RSD&ASR$\uparrow$&\makecell{Avg\\(RoI-L)}&\%RSD$\downarrow$ & ASR$\uparrow$&\makecell{Avg\\(RoI-L)}&\%RSD$\downarrow$&ASR$\uparrow$&\makecell{Avg\\(RoI-L)}&\%RSD$\downarrow$ \\
\midrule
\multirow{4}{*}{\makecell{RTX 2080\\Super}} 
& AttFusion & 0.039 & 53.85\% & 1.27\% & \makecell{0.806\\(23.8$\times$)} & 31.89\% & 18.82\% & \makecell{1.076\\(26.9$\times$)} & 50.37\% & \textbf{99.61\%} & \makecell{\textbf{2.981}\\(\textbf{89.4$\times$})} & \textbf{7.98\%} \\
& CoAlign   & 0.050 & 48.00\% & 50.98\% & \makecell{1.646\\(33.5$\times$)} & 37.85\% & 76.44\% & \makecell{2.364\\(47.1$\times$)} & 44.46\% & \textbf{100\%} & \makecell{\textbf{4.078}\\(\textbf{91.3$\times$})} & \textbf{4.39\%} \\
& V2VAM     & 0.042 & 38.10\% & 0.00\% & \makecell{0.090\\(1.3$\times$)} & 113.33\% & 0.00\% & \makecell{0.302\\(6.0$\times$)} & 55.30\% & \textbf{15.26\%} & \makecell{\textbf{1.011}\\(\textbf{22.8$\times$})} & \textbf{48.86\%} \\
& W2C& 0.051 & 33.33\% & 0.00\% & \makecell{0.041\\($-$0.2$\times$)} & 24.39\% & 0.00\% & \makecell{0.070\\(0.3$\times$)} & 55.71\% & \textbf{96.93\%} & \makecell{\textbf{2.251}\\(\textbf{46.5$\times$})} & \textbf{13.95\%} \\
\midrule
\multirow{4}{*}{RTX 3060 Ti} 
& AttFusion & 0.033 & 106.06\% & 0.78\% & \makecell{0.667\\(22.2$\times$)} & 61.02\% & 7.74\% & \makecell{0.797\\(22.8$\times$)} & 48.81\% & \textbf{99.15\%} & \makecell{\textbf{2.164}\\(\textbf{64.6$\times$})} & \textbf{8.78\%} \\
& CoAlign   & 0.037 & 127.03\% & 6.36\% & \makecell{1.043\\(28.4$\times$)} & 27.33\% & 38.83\% & \makecell{1.374\\(36.7$\times$)} & 42.94\% & \textbf{100\%} & \makecell{\textbf{2.898}\\(\textbf{69.7$\times$})} & \textbf{5.24\%} \\
& V2VAM     & 0.034 & 64.71\% & 0.00\% & \makecell{0.065\\(1.0$\times$)} & 90.77\% & 0.00\% & \makecell{0.149\\(3.3$\times$)} & 71.81\% & \textbf{0.33\%} & \makecell{\textbf{0.636}\\(\textbf{17.2$\times$})} & \textbf{49.69\%} \\
& W2C& 0.045 & 62.22\% & 0.00\% & \makecell{0.035\\($-$0.2$\times$)} & 5.71\% & 0.00\% & \makecell{0.044\\(0.0$\times$)} & 13.64\% & \textbf{92.38\%} & \makecell{\textbf{1.754}\\(\textbf{33.3$\times$})} & \textbf{10.95\%} \\
\midrule
\multirow{4}{*}{RTX 4090} 
& AttFusion & 0.018 & 72.22\% & 0.00\% & \makecell{0.382\\(24.6$\times$)} & 30.89\% & 0.00\% & \makecell{0.483\\(27.2$\times$)} & 49.48\% & \textbf{1.07\%} & \makecell{\textbf{1.290}\\(\textbf{81.1$\times$})} & \textbf{8.06\%} \\
& CoAlign   & 0.021 & 76.19\% & 1.07\% & \makecell{0.719\\(34.9$\times$)} & 35.88\% & 14.24\% & \makecell{1.000\\(47.4$\times$)} & 45.60\% & \textbf{99.76\%} & \makecell{\textbf{1.728}\\(\textbf{86.9$\times$})} & \textbf{4.51\%} \\
& V2VAM     & 0.023 & 69.57\% & 0.00\% & \makecell{0.035\\(0.8$\times$)} & 114.29\% & 0.00\% & \makecell{0.142\\(6.0$\times$)} & 58.45\% & 0.00\% & \makecell{\textbf{0.478}\\(\textbf{23.5$\times$})} & \textbf{47.70\%} \\
& W2C& 0.026 & 46.15\% & 0.00\% & \makecell{0.015\\($-$0.4$\times$)} & 26.67\% & 0.00\% & \makecell{0.031\\(0.2$\times$)} & 80.65\% & \textbf{0.59\%} & \makecell{\textbf{1.040}\\(\textbf{48.3$\times$})} & \textbf{19.62\%} \\
\bottomrule
\end{tabular}
\caption{Results of average latency (s), relative standard deviation ($\%$RSD), rate-of-increase in latency (RoI-L), and attack success rate (ASR) under 1.5\,s threshold. \attack significantly increases latency more reliably than PGD~\cite{MadryMSTV17PGD} and prior art~\cite{liu2023slowlidar, ma2024slowtrack}.}
\label{tab:latency_results}
\end{table*}

\subsubsection{Implementation Details.}
We use the public implementations and pre-trained weights of the four CP algorithms released for OPV2V~\cite{xu2022opv2v}. For \attack, we set \(\lambda_1=0.1\) and \(\lambda_2=1\). We employ the BIM optimization scheme~\cite{kurakin2018bim} with \(10\) gradient descent steps, each having a step size of \(0.1\). The same optimization parameters are applied to both baseline attacks. 
All model hyperparameters follow the official default configuration~\cite{opencood}.

\subsection{Evaluation Results}\label{sec:eval_results}

\subsubsection{Proposal Count Inflation.} 
Table~\ref{tab:roi_proposal} shows how \attack sharply increases the average number of bounding box proposals \(|\mathcal{P}|\) across all four CP algorithms. In contrast, PGD and prior art induce only moderate or even negative (\(-0.62\times\)) growth for certain models (\eg W2C). For instance, PGD induces an average $\text{RoI-P}$ of \(12.28\times\), and prior art of \(13.26\times\), whereas \attack achieves a notably higher \(46.19\times\). Similar proposal count increase patterns are observed on the other CP models.

\subsubsection{Latency across Hardware.} 
Figure~\ref{fig:latency_boxplot} illustrates the distribution of the end-to-end latency of the victim’s CP pipeline across different hardware.  
Table~\ref{tab:latency_results} summarizes the latency of the victim’s CP pipeline and compares across different attack methods. We observe that PGD rarely exceeds a \(1.3\text{-}2.0\times\) overhead on algorithms like V2VAM and W2C, sometimes failing to increase latency at all (\eg W2C on 2080 Super yields \(-0.2\times\)). Prior art achieves moderate increases, but remains below the dramatic slowdowns caused by \attack. Higher computing power, as shown by the RTX 4090, reduces the latency effect, yet \attack consistently yields the highest RoIs in latency--often surpassing \({80\times}\), with the lowest $\%\text{RSD}$.

\subsubsection{Attack Success under Different Latency Thresholds.} 
Figure~\ref{fig:asr} shows the ASR as a function of the latency threshold, which represents different levels of real-time responsiveness required by the system. 
\attack achieves higher ASRs across different thresholds and hardware than the prior art baseline. 
For example, on RTX 2080 Super, \attack can push the latency of V2VAM beyond 1 s with around 40\% ASR, on which the prior art can only introduce a latency less than 0.1 s with less than 20\% ASR. Even at more strict thresholds around 2 s, \attack maintains an ASR above 80--100\% for the other models.  

\subsection{Ablation Study}
\label{sec:ablation}

We further investigate the effectiveness of \attack under more restrictive hyperparameter settings in CP post-processing. Specifically, we vary the confidence threshold (for filtering out low-confidence pre-NMS proposals), the IoU threshold in NMS, and the maximum number of proposals allowed into NMS. By default, these values are set to \((0.2,\,0.15,\,1000)\) respectively~\cite{opencood}.

\subsubsection{Confidence Threshold.}
\begin{figure}[t]
    \begin{tabular}[t]{c}
    \centering
    \begin{subfigure}[t]{0.48\columnwidth}
        \includegraphics[width=\linewidth]{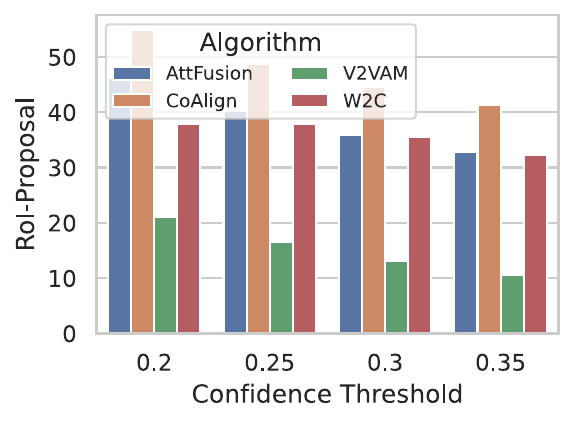}
        \caption{RoI-P}
        \label{fig:roi_proposal_conf}
    \end{subfigure}
    \begin{subfigure}[t]{0.48\columnwidth}
        \includegraphics[width=\linewidth]{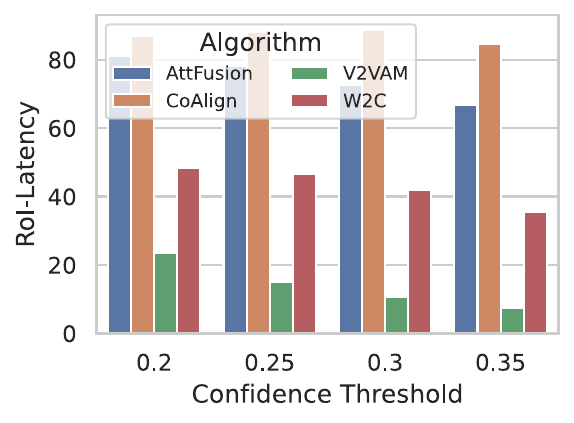}
        \caption{RoI-L (RTX 4090)}
        \label{fig:roi_latency_conf}
    \end{subfigure}
\end{tabular}
	\caption{Attack effectiveness under different pre-NMS confidence thresholds. \attack maintains high proposal count increase and slow-down effects for higher confidence thresholds. }
\label{fig:ablation_conf}
\end{figure}

Figure~\ref{fig:ablation_conf} shows that raising the confidence threshold only marginally reduces the RoI-P and RoI-L. Higher thresholds admit fewer borderline proposals, thereby reducing NMS overhead. However, even at higher thresholds (\eg 0.35), \attack still causes detection to be tens of times slower, indicating that simply tightening this threshold cannot neutralize the threat. 

\subsubsection{IoU Threshold and Maximum Number of Proposals.}
\begin{figure}[t]
    \begin{tabular}[t]{c}
    \centering
    \begin{subfigure}[t]{0.48\columnwidth}
        \includegraphics[width=\linewidth]{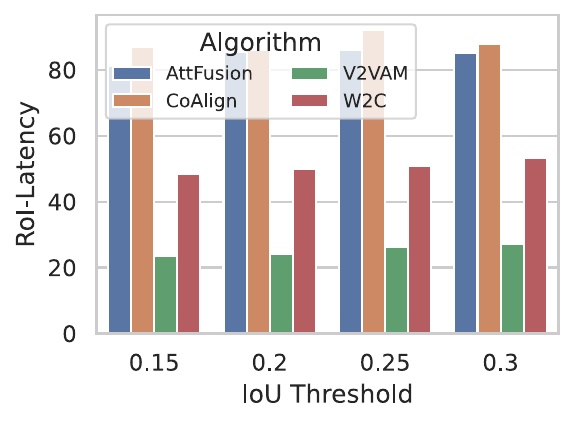}
        \caption{RoI-L vs IoU Threshold}
        \label{fig:roi_latency_iou}
    \end{subfigure}
    \begin{subfigure}[t]{0.48\columnwidth}
        \includegraphics[width=\linewidth]{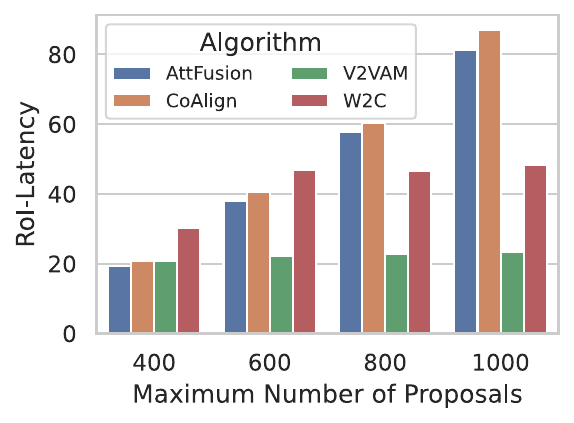}
        \caption{RoI-L vs Max. \#Proposals}
        \label{fig:top_k}
    \end{subfigure}
\end{tabular}
\caption{RoI-L on RTX 4090 for different NMS hyperparameters. \attack achieves similar slowdown effects across different IoU thresholds, indicating high persistence of the fabricated proposals. Lowering the threshold for the maximum number of proposals can only partially alleviate the latency effect.}\label{fig:ablation_latency}
\end{figure}

Raising the IoU threshold allows inflated boxes to persist longer, as fewer proposals are eliminated per iteration. As shown in Figure~\ref{fig:roi_latency_iou}, the similar slowdown effect at various IoU thresholds achieved by \attack reflects the high persistence of the small and distributed proposals created. 
Figure~\ref{fig:top_k} shows that capping the number of proposals below 1,000 partially alleviates the slowdown but still leaves a substantial latency increase (up to \(20\times\)). 

\subsubsection{Spatial-Temporal Warping.}
Table~\ref{tab:warping_ablation} highlights the impact of our spatial-temporal feature warping on the performance of \attack. In most cases, warping significantly increases both RoI-P and RoI-L on RTX 4090, illustrating how aligning feature maps across frames, despite asynchronous knowledge, effectively creates more adversarial proposals, hence slowing down the victim's CP system.

\begin{table}[t]\centering
\footnotesize
\begin{tabular}{l|r|r|r|rr}\toprule
Metric &\multicolumn{2}{c|}{RoI-P$\uparrow$} &\multicolumn{2}{c}{RoI-L$\uparrow$} \\\cmidrule{1-5}
Model &w/o warp &w/ warp &w/o warp &w/ warp \\\midrule
AttFusion &26.14$\times$ &\textbf{46.19$\times$} &63.91$\times$ &\textbf{81.10$\times$} \\
CoAlign &30.19$\times$ &\textbf{54.84$\times$} &\textbf{98.74$\times$} &86.90$\times$ \\
V2VAM &9.97$\times$ &\textbf{21.09$\times$} &6.35$\times$ &\textbf{23.50$\times$} \\
W2C &3.47$\times$ &\textbf{37.81$\times$} &0.70$\times$ &\textbf{48.30$\times$} \\
\bottomrule
\end{tabular}
\caption{Spatial-temporal warping increases attack effectiveness with higher RoIs in both proposal counts and latency.}\label{tab:warping_ablation}
\end{table}

\subsection{Attack Robustness Against Defenses}
\label{sec:defense_eval}

\subsubsection{Adversarial Training.} We retrained models for five epochs while exposing them to \attack. This defense was ineffective. For most models, it caused catastrophic collapse (0 mAP), destroying their utility. For V2VAM, which remained functional (0.89 mAP), \attack still achieved a potent $18.7\times$ RoI-L. This failure occurs because the objectives of integrity robustness and computational efficiency are orthogonal. Training against latency attacks does not teach the model to be faster, but instead can corrupt its learned feature representations.

\subsubsection{ROBOSAC.} We tested \attack against ROBOSAC~\cite{amongUs}, a SOTA integrity defense that uses a RANSAC-like approach to identify malicious agents. As Table~\ref{tab:robosac} shows, ROBOSAC restores detection integrity (mAP). However, its iterative design is fundamentally incompatible with defending against latency attacks. By repeatedly running the already-slowed perception pipeline, ROBOSAC inadvertently \emph{amplifies} the latency, increasing the RoI-L from $81.1\times$ to $777.8\times$ for AttFusion. This demonstrates that reactive defenses designed for integrity can exacerbate availability threats.

\begin{table}[t]\centering
\footnotesize
\begin{tabular}{l|c|c|c|cr}\toprule
Metric &\multicolumn{2}{c|}{mAP@0.5$\uparrow$} &\multicolumn{2}{c}{RoI-L$\uparrow$} \\\cmidrule{1-5}
Model & \begin{tabular}[c]{@{}c@{}}No\\Defense\end{tabular} & ROBOSAC & \begin{tabular}[c]{@{}c@{}}No\\Defense\end{tabular} & ROBOSAC \\\midrule
AttFusion & 0 & \textbf{0.71} & 81.10$\times$ & \textbf{777.85$\times$} \\
CoAlign & 0 & \textbf{0.72} & 86.90$\times$ & \textbf{785.54$\times$} \\
V2VAM & 0 & \textbf{0.77} & 23.50$\times$ & \textbf{221.44$\times$} \\
W2C & 0 & \textbf{0.74} & 48.30$\times$ & \textbf{451.93$\times$} \\
\bottomrule
\end{tabular}
\caption{ROBOSAC guards the integrity of cooperative perception, restoring mAP, yet amplifies the latency effect caused by \attack.}\label{tab:robosac}
\end{table}


\section{Discussion}
\label{sec:discussion}

\subsubsection{Practical Impact.}
Our work presents the first latency attack on CP, wherein a practical attacker can inflate the victim’s perception latency effectively by injecting adversarial perturbations into the shared messages. On our real-world vehicle testbed, the attacker can generate online perturbations in real-time with 50\,ms per iteration, demonstrating its practical feasibility. We further confirm the applicability of \attack on an NVIDIA Jetson Orin SoC platform. Although the public benchmark CP implementation~\cite{opencood} runs more slowly on Orin at around 0.3\,s per benign frame (potentially due to lack of official software hardware-co-design), \attack nonetheless achieves similar Rate-of-Increase in latency, highlighting the hardware-agnostic nature of our approach. Since a single compromised CAV can degrade perception for multiple vehicles, our results highlight an urgent need for more robust CP system designs.

\subsubsection{Potential Defenses.}
Current CP defenses~\cite{amongUs, zhang2024datafab} require available final detection results to verify or repair, which makes them susceptible to \attack. Tuning hyperparameters can reduce the number of proposals allowed and cap the worst-case scenario, but at the cost of a higher risk of overlooking true detections under benign cases and an inability to fully mitigate the slowdown effect. Therefore, more proactive defenses are needed to counter the availability threats posed by \attack.

\subsubsection{Black-box Attack.}
Our attack assumes white-box knowledge of the CP pipeline, in line with existing work~\cite{liu2023slowlidar, ma2024slowtrack}. However, emerging heterogeneous CP systems~\cite{lu2024heal} create a new system model where collaborators do not use the same CP algorithm. This translates into a black-box attack setting. 
As future work, we plan on investigating query-based black-box attacks, where the adversary only observes its own CP algorithm output and timing.\looseness=-1


\section{Conclusion}
\label{sec:conclusion}
We present \attack, the first latency attack against vehicular cooperative perception (CP) by injecting adversarial perturbations via V2X messages, threatening the availability and timeliness guarantees of the safety-critical autonomous driving systems. 
Through extensive experiments on the benchmark dataset OPV2V with various SOTA CP algorithms and hardware platforms, \attack shows slowdowns of $90\times$ over benign operation, causing the single-frame latency to exceed 3 s with 100\% success rates on our vehicle testbed. 
Our work highlights the need for new defenses to counter such availability threats.

\bibliography{aaai2026}

\end{document}